\documentclass[prd,amsmath,amssymb,reprint]{revtex4-2}
\usepackage{graphicx} 
\usepackage{dcolumn}  
\usepackage{bm}       
\usepackage{hyperref} 
\usepackage[utf8]{inputenc} 
\usepackage{siunitx}  
\usepackage{graphicx}
\usepackage{tikz}
\usepackage{natbib}
\usepackage{braket}
\usepackage{subcaption}

\begin{document}

\title{The effects of Gravitational Field on the Efimov State}

\author{Feng Yixuan}
\email{yf1025@ic.ac.uk}
\affiliation{Department of Physics, Imperial College London, South Kensington, London, SW7 2AZ}

\begin{abstract}

In hyperspherical coordinates, by incorporating the boundary conditions of the short-range potential, we successfully reproduced the derivation of Efimov energy levels in three-dimensional space. In subsequent discussions, we extended this derivation to Schwarzschild spacetime. By combining the Schrödinger equation in curved spacetime, we obtained the corrections to the Efimov energy spectrum under weak gravitational fields. Furthermore, using Riemann normal coordinate expansions, we rigorously derived the hyperradial equation in the presence of strong gravitational fields and applied first-order perturbation theory to compute the internal energy corrections of the Efimov system under strong-field conditions. We further propose a potential experimental approach to validate the theory by simulating spacetime curvature effects induced by gravity through the use of ultra-cold atoms in optical lattices under laboratory conditions.

\end{abstract}

\maketitle

\section{Introduction}

In 1970, Efimov theoretically predicted that the three particles (nucleons or atoms) can form infinite bound states with energy level $E_n=E_0e^{-2n\pi/s_0}$ when the scattering length of any one pair of particles approaches infinity ($a\to\infty$) and the interaction potential is a short-range potential. \cite{1973}

In 2006, research teams from the University of Innsbruck and the University of Chicago successfully observed Efimov states for the first time using ultracold caesium atoms \cite{2006}, where the scattering length was tuned via Feshbach resonance. In subsequent years, numerous experiments in cold atomic systems (e.g., lithium \cite{Li}, potassium) and nuclear physics (e.g., tritium, helium isotopes \cite{He4}) have confirmed the existence of this phenomenon.

Efimov physics also plays a crucial role in stellar nucleosynthesis, for example, the Hoyle state of carbon-12, which is a crucial excited nuclear state (with quantum numbers $0^+$.) that plays a key role in the triple-$\alpha$ process, allowing the formation of carbon in stars. It is predicted by Fred Hoyle (1954) to explain the carbon abundance in the universe. \cite{C12} Since the $\alpha$ particles are electrically charged, they will be subjected to the Coulomb interaction, which is a long-range force whose potential decays as $1/r$. Therefore, the Efimov states can be maintained by the Efimov attractive potential at short distances ($V\sim1/R^2$), but they may vanish at long ranges due to the presence of the Coulomb potential ($V\sim1/r$). Hence, the Hoyle state is suggested as a resonant state resulting from the balance between the Efimov attractive force and the Coulomb repulsive force.\cite{colomb1},\cite{colomb2}

The Efimov states typically form under low-energy conditions where relativistic corrections are generally negligible. However, when the interaction range between particles becomes smaller than the Compton wavelength, relativistic effects must be taken into account. This scenario was discussed in the 1980s by James V. Lindesay and H. Pierre Noyer. \cite{re1},\cite{re2}

In 2018, an investigation attempted to extend Efimov physics to curved spacetime \cite{GeneralEfimov}. However, the study only qualitatively analysed the potential influence of gravitational fields on Efimov physics, without providing a rigorous mathematical treatment.

In the 2020 study, it was discovered that the solutions to the Schrödinger equation on hyperbolic surfaces exhibit discrete scale symmetry analogous to Efimov states. Moreover, this scheme can be generalised to higher-dimensional surfaces or spaces with complex metrics, providing new tools for quantum simulation and gravitational theory \cite{hyperbolic}, but the paper does not discuss Efimov states in gravitational fields from the perspective of general relativity.

This work theoretically extends the study of Efimov states to Schwarzschild spacetime and predicts gravitational corrections to Efimov energy levels. We expect these results to offer valuable insights for quantum gravity research. The proposed methodology for analysing Efimov states in curved spacetime bridges a critical gap in Efimov physics within gravitational backgrounds.

\section{Efimov State in Flat Space}

In this section, the key techniques and procedures needed to derive the Efimov state are outlined. The established formalism provides a foundation for extending the analysis to curved spacetime.

\subsection{Jacobi Coordinate}

Consider a quantum system formed by three particles, for instance, two up quarks and one down quark in a proton. There is an interaction between three particles that requires a non-separable treatment of the wavefunction. Therefore, a fully correlated wavefunction must be employed to accurately describe the system.

The wavefunction of the system can be written as:

\begin{equation}
    \hat{H}\Psi=\sum_{i=1}^3\frac{\hbar^2}{2m_i}\nabla^2_i\Psi+\sum_{i>j}V_{ij}(r_{ij})\Psi=E\Psi
\end{equation}

Assume the positions of the three particles are \( \vec{r}_1, \vec{r}_2, \vec{r}_3 \), with corresponding masses \( m_1, m_2, m_3 \).

\textbf{Jacobi relative coordinates} are defined as:

\begin{equation}
    \vec{X} = \vec{x}_2 - \vec{x}_1
\end{equation}

(relative position of particles 1 and 2)

\begin{equation}
    \vec{Y} = \vec{x}_3 - \frac{m_1 \vec{x}_1 + m_2 \vec{x}_2}{m_1 + m_2}
\end{equation}

(position of particle 3 relative to the centre of mass of particles 1 and 2)

\begin{equation}
\vec{r}_0 =\frac{m_1 \vec{x}_1 + m_2 \vec{x}_2+m_3\vec{x}_3}{m_1 + m_2+m_3}
\end{equation}

(Position of the centre mass of the three-body system.)

These two variables ($X$ and $Y$) are the \textbf{internal coordinates} of the system (they do not include centre-of-mass motion).

The three-body kinetic energy is originally expressed as:

\begin{equation}
    \hat T = - \sum_{i=1}^3 \frac{\hbar^2}{2m_i} \nabla_i^2
\end{equation}

Change of variables is used in Jacobi coordinates and apply an appropriate scaling (mass-weighted) so that the kinetic energy becomes (The detailed derivation is provided in the Appendix \ref{B}):

\begin{equation}
    \hat T = -\frac{\hbar^2}{2\mu_X} \nabla_{X}^2 
    - \frac{\hbar^2}{2\mu_Y} \nabla_{Y}^2 
    - \frac{\hbar^2}{2M} \nabla_{\text{cm}}^2
\end{equation}

where:

\begin{itemize}
  \item $\mu_X = \dfrac{m_1 m_2}{m_1 + m_2}$: reduced mass of particles 1 and 2.
  \item $\mu_Y =\dfrac{m_3 (m_1 + m_2)}{m_1 + m_2 + m_3}$: reduced mass of particle 3 relative to the centre of mass of particles 1 and 2.
  \item $M = m_1 + m_2 + m_3$: total mass of the system.
\end{itemize}

In the flat spacetime, the motion of the CoM (centre of mass) point is at constant velocity since there is no external force acting on the system. Therefore, we can study the system in the CoM frame to eliminate 3 degrees of freedom to simplify the problem.

In other words, the kinetic operator is only related to the internal structure of the system.

\begin{eqnarray}
    \hat{T}=\hat{T}_{\text{Internal}}=-\frac{\hbar^2}{2\mu_X} \nabla_{X}^2 
    - \frac{\hbar^2}{2\mu_Y} \nabla_{Y}^2
\end{eqnarray}

\subsection{Hyperspherical Coordinate}

To obtain the expression for the internal energy, we shall rewrite the kinetic operator in one term. We first define that $\vec{\rho} =\vec{X}\sqrt{\mu_X/m_0}$ and $\vec{\lambda} =\vec{Y}\sqrt{\mu_Y/m_0}$, where $m_0$ is the unit mass. By doing this, the kinetic operator can be rescaled into: $\hat{T}=-\frac{\hbar^2}{2m_0}\left(\nabla^2_\rho+\nabla^2_\lambda\right)$.\\

The \textbf{hyperradius} and \textbf{hyperangles} are defined as \cite{chi}:

\begin{equation}
\begin{cases}
\zeta = \sqrt{ \mathbf{\rho}^2 + \mathbf{\lambda}^2}, \qquad \text{with } \zeta \in \left[\zeta_{\text{min}}, \zeta_{\text{max}} \right]\\[2.5ex]
\alpha = \arctan\left( \frac{|\mathbf{\rho}|}{|\mathbf{\lambda}|} \right), \quad \text{with } \alpha \in \left[0, \frac{\pi}{2} \right]
\end{cases}
\end{equation}

Along with the angular directions of \( \vec{\rho} \) and \( \vec{\lambda} \), the full set of internal variables consists of:
\[
(\zeta, \alpha, \hat{\boldsymbol{\rho}}, \hat{\boldsymbol{\lambda}}) \in \mathbb{R}^6
\]

\vspace{1em}

The internal kinetic operator may be written as:

\begin{equation}
\hat T_{\text{internal}} = -\frac{\hbar^2}{2m_0} \left( \frac{\partial^2}{\partial \zeta^2} + \frac{5}{\zeta} \frac{\partial}{\partial \zeta} - \frac{\hat\Lambda^2}{\zeta^2} \right)\label{Tinternal}
\end{equation}

Where $\hat\Lambda^2$ is the hyperangular operator.

Note that while the hyperradius $\zeta$ theoretically spans from 0 to infinity, its practical range is physically constrained by the scattering length, as will be discussed in detail later.

\subsection{Bethe-Peierls boundary condition}

The Efimov state can form only when the interaction potential is a short-range potential, which means the potential can be treated as a delta function potential, e.g, $V(\vec r)\sim\delta(\vec r)$. This means that we can consider the system as free (no interaction) and treat the potential as a boundary condition; this boundary condition is called the Bethe-Peierls boundary condition.\cite{bethe}

\begin{equation}
    \lim_{r\to 0}\frac{\partial}{\partial r}\left[r\psi(r)\right]=-\frac{1}{a}\lim_{r\to 0}[r\psi(r)]
\end{equation}

\subsection{Energy level structure}

The three particles are all bosons because of the bosonic exchange symmetry; the wavefunction can be decomposed as follows \cite{chi}:

\begin{equation}
\Psi = \chi(\vec{\rho}_{12}, \vec{\lambda}_{12,3}) + \chi(\vec{\rho}_{23}, \vec{\lambda}_{23,1}) + \chi(\vec{\rho}_{31}, \vec{\lambda}_{31,2})
\end{equation}

Where $\chi$ is known as the Faddeev component. Assume that three bosons have the same mass, $\vec{\rho}_{12}=\frac{\sqrt{2}}{2}\left(\vec{r}_2-\vec{r}_1\right)$ and $\vec{\lambda}_{12,3}=\sqrt{\frac{2}{3}}\left(r_3-\frac{1}{2}(r_1+r_2)\right)$. For the other two situations, their Jacobi coordinates can be expressed in the basis $(\vec\rho_{12},\vec\lambda_{12,3})$.
Notice that the system only contains Bosons. They can stay in the same quantum state. For the $s$ wave, the solution is symmetric. This means that the solution is located in a 2D space instead of a 6D space, e.g $(\zeta, \alpha) \in \mathbb{R}^2$. Express the kinetic energy operator in Eq. \ref{Tinternal} into the Jacob coordinate system:

\begin{eqnarray}
    \hat T_{\text{internal}}&=&-\frac{\hbar^2}{2m_0} \left( \nabla_{\rho}^2 + \nabla_{\lambda}^2 \right)\\&=&-\frac{\hbar^2}{2m_0} \left( \frac{\partial^2}{\partial \zeta^2} + \frac{1}{\zeta} \frac{\partial}{\partial \zeta} + \frac{1}{\zeta^2}\frac{\partial^2}{\partial\alpha^2} \right)
\end{eqnarray}

The wavefunctions can be written as $\chi(\rho,\lambda)=\frac{\chi_0(\rho,\lambda)}{\rho\lambda}$ \cite{Efimov}, applying the Bethe-Peierls boundary condition to the system for $\rho$ as figure 1 shows:

\begin{equation}
    \lim_{\rho\to0}\frac{\partial}{\partial\lambda}\left(\chi_0(\rho,\lambda)\right)+\frac{\chi_0\left(\frac{\sqrt3}{2}\lambda,\frac{1}{2}\lambda\right)}{\frac{\sqrt3}{4}\lambda^2}+\frac{\chi_0\left(\frac{\sqrt3}{2}\lambda,\frac{1}{2}\lambda\right)}{\frac{\sqrt3}{4}\lambda^2}=0
\end{equation}

\begin{figure}[htbp]
    \centering
    \begin{subfigure}{0.46\textwidth} 
        \centering
        \includegraphics[width=\textwidth]{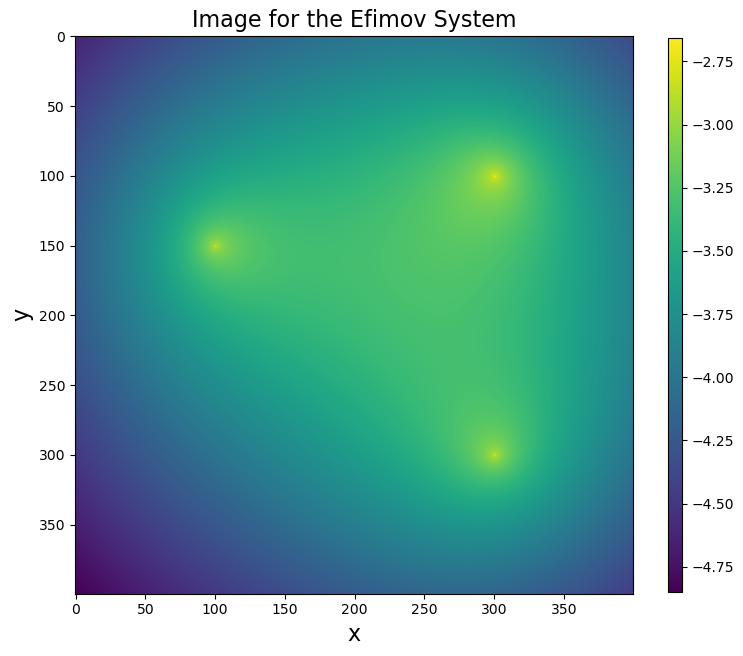}
        \caption{Three Bosons at low energy ($s$ wave.)}\label{fig:subimage1}
    \end{subfigure}
    \hspace{3pt} 
    \begin{subfigure}{0.46\textwidth} 
        \centering
        \includegraphics[width=\textwidth]{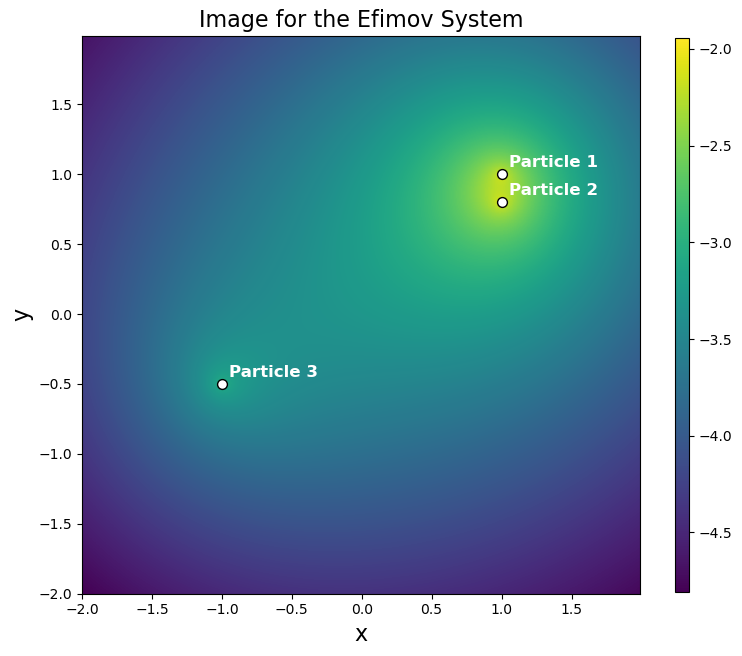}
        \caption{This is the limit case for $\rho$ approach to zero.}
        \label{fig:subimage3}
    \end{subfigure}

    \caption{Efimov states are an infinite series of three-body bound states that can arise when the two-body subsystems are close to binding (near-resonant), even if each pair cannot bind deeply.}
    \label{fig:combined_figure}
\end{figure}

Expressing the equation above in the hyperspherical coordinate at the $\alpha\to0$ limit:

\begin{equation}
    \lim_{\alpha\to0}\frac{\partial}{\partial\alpha}\left(\chi_0(\zeta,\alpha)\right)+\frac{8}{\sqrt3}\chi_0\left(\zeta,\frac{\pi}{3}\right)=0\label{15}
\end{equation}

By separating variables $\chi_0(\zeta,\alpha)=F(\zeta)\Phi(\alpha)$ and solving the hyperangular differential equation, we can get $\Phi(\alpha)\sim\sin(s_n(\alpha-\pi/2))$, take it back to the Eq. \ref{15}:

\begin{equation}
    - s_n \cos\left( \frac{\pi s_n}{2} \right) 
+ \frac{8}{\sqrt{3}} \sin\left( \frac{\pi s_n}{6} \right) = 0
\end{equation}

There will be only one imaginary solution $|s_0|\approx 1.00624$.\\

The hyper-radial equation is:

\begin{equation}
    F''(\zeta)+\frac{1}{\zeta}F'(\zeta)+\frac{|s_0|^2}{\zeta^2}F(\zeta)=-\frac{2m_0E}{\hbar^2}F(\zeta)
\end{equation}
The strict solution to this equation is:

\begin{equation}
    F(\zeta)\sim K_{is_0}\left(\frac{\sqrt{2m_0E}}{\hbar}\zeta\right)
\end{equation}

Where $K_{a}(kx)$ denotes the modified Bessel function of the second kind. At the low energy limit, the solution is:

\begin{equation}
    F(\zeta) \sim\sin\left( |s_0| \ln \zeta+\delta\right)
\end{equation}

Applying the cutoff condition:
\begin{equation}
    F(\zeta_0) \sim\sin\left( |s_0| \ln \zeta_0+\delta\right)=0
\end{equation}
Therefore, $|s_0| \ln \zeta_0+\delta=\pm n\pi$. The phase difference can be written as:

\begin{equation}
    \delta_n=-|s_0| \ln \zeta_0\pm n\pi=-|s_0|\ln(\zeta_0e^{\mp n\pi/|s_0|})
\end{equation}

\begin{equation}
    F_n(\zeta) \sim\sin\left( |s_0| \ln \zeta+\delta_n\right)=\sin\left( |s_0| \ln\left[ \zeta /( \zeta_0e^{n\pi/s_0})\right] \right)\label{Fnzeta}
\end{equation}

$\zeta_0e^{n\pi/|s_0|}$ is chosen rather than $\zeta_0e^{-n\pi/|s_0|}$ since $\zeta_0$ is the short-range cutoff.

Hence, it is convenient to choose a periodic window to study relevant physical properties. The principle can be analogised to the standing wave equation solution in an infinite potential well. The energy level is found:

\begin{equation}
    E_n=-\frac{\hbar^2}{2m\zeta_n^2}=E_0e^{-2n\pi/s_0}
\end{equation}\\
\section{Efimov State in Curved Space (Schwarzschild spacetime)}
In this section, the Efimov energy level change under the influence of the gravitational field is considered. To extend the Efimov theory to curved spacetime,  we shall start from the basic example.

\begin{tikzpicture}[scale=1.5]

    \filldraw[black] (0,0) circle (0.1cm);
    \filldraw[black!70] (0,0) circle (0.0cm);
        
    \filldraw[blue] (2,-2.5) circle (0.05cm) node[below=0.2cm] {Observer};
    \filldraw[blue] (4.0,-0.5) circle (0.03cm) node[below=0.2cm] {Partice 1 ($\vec r_1$)};
    \filldraw[blue] (3.5,-0.1) circle (0.03cm) node[above=0.2cm] {Particle 2 ($\vec r_2$)};
    \filldraw[blue] (3,-1.5) circle (0.03cm) node[below=0.2cm] {Particle 3 ($\vec r_3$)};
        
    \draw[red, ->, >=stealth] (0,0) -- (4.0,-0.5);
    \draw[red, ->, >=stealth] (0,0) -- (3.5,-0.1);
    \draw[red, ->, >=stealth] (0,0) -- (3,-1.5);

    \draw[black, -, >=stealth] (4.0,-0.5) -- (3.5,-0.1);
    \draw[black, -, >=stealth] (4.0,-0.5) -- (3,-1.5);
    \draw[black, -, >=stealth] (3,-1.5) -- (3.5,-0.1);
        
    \node at (0,-0.8) {Black Hole};
    \node at (3.4,-2.3) {Efimov System};
    \node at (2,-3.3) {\textbf{Diagram 1}};
\end{tikzpicture}

As Diagram 1 shows, assuming the Efimov system near a blackhole (Schwarzschild spacetime), one observer on Earth (Minkowski spacetime). The metrics are given by equations below:

{\begin{eqnarray}
    ds^2&=&g_{\mu\nu}dx^\mu dx^\nu \nonumber\\
      &=&-f(r) d\tau^2 + \frac{1}{f(r)} dr^2+r^2d\Omega^2\label{24}
\end{eqnarray}}

Where $f(r)=1-\frac{2M}{r}$, $d\Omega^2=d\theta^2 + \sin^2\theta \, d\phi^2$ and:

{\begin{eqnarray}
    ds^2=\eta_{\mu\nu}dx^\mu dx^\nu=-dt^2 + dr^2+r^2 d\Omega^2\label{25}
\end{eqnarray}}

It is worth noting that the internal scale of the three-body quantum system is much smaller than the distance from the centre of mass of the three-body system to the black hole. Therefore,  the gravitational influence on Efimov states can be discussed in two separate cases: \textbf{weak field} and \textbf{strong field}.

\subsection{Efimov State in Weak Field}\label{Weakfield}

In this part, an Efimov system is placed near a massive blackhole. \textbf{The weak field is defined as: The gravitational field is not strong enough to alter the space structure in the internal system.} In other words, spacetime in the system is still Minkowski spacetime. Therefore, the structure of the Efimov state stays the same. However, although the system structure will not change, the centre of mass energy will be redshifted because of the blackhole.\\
For the observer on Earth, we noticed that:
\begin{equation}
i\hbar \frac{\partial \Psi}{\partial \tau} = -\frac{\hbar^2}{2m} \nabla_g^2 \Psi\label{26}
\end{equation}

Where $\nabla_g^2$ is \cite{GRquantum}:

\begin{equation}
\nabla_g^2 \Psi = \frac{1}{\sqrt{|g|}} \partial_i \left( \sqrt{|g|} g^{ij} \partial_j \Psi \right)
\end{equation}

The equation above can be shown via tensor analysis. In Schwarzschild spacetime, the metric is given by Eq. \ref{24} ($\hbar=c=G=1$), but for the particle itself, the spacetime is locally flat, so the metric is Minkowski as equation \ref{25} shows.

The left-hand side of the Eq. \ref{26} can be written in locally flat time $t$:

\begin{equation}
    \hat{H}=i\hbar\frac{\partial}{\partial \tau}=i\hbar\frac{\partial t}{\partial \tau}\frac{\partial}{\partial t}
\end{equation}

According to the metric provided in Eqs. \ref{24} and \ref{25}:

\begin{equation}
    \frac{\partial t}{\partial \tau}=\sqrt{f(r)}=\sqrt{1-\frac{2M}{r}}
\end{equation}

\begin{equation}
    \hat{H}=i\hbar\frac{\partial}{\partial \tau}=\sqrt{1-\frac{2M}{r}}\hat{H_0}
\end{equation}

Therefore, the energy observed on Earth is:

\begin{equation}
    E=\sqrt{1-\frac{2M}{r}}E_0
\end{equation}

Where $E_0$ is the energy of the particle in the flat space.

The equation above shows that the quantum system energy will be redshifted because of the gravitational field. Noticed that in the Jacob coordinate system, the Hamiltonian operator for the Efimov state can be written as:

\begin{equation}
\hat{H}=\hat{H}_{\text{cm}}+\hat{H}_{\text{In}}
\end{equation}

The internal energy stays the same:

\begin{equation}
\hat{H}_{\text{In}}\psi_{\text{In}}=E^{\text{In}}_0e^{-2n\pi/s_0}\psi_\text{In}
\end{equation}

The center of mass energy is:

\begin{equation}
    \hat{H}_{\text{cm}}\psi_{\text{cm}}=\sqrt{1-\frac{2M}{r}}E^{\text{cm}}_0\psi_\text{cm}
\end{equation}

Therefore, the total energy is:

\begin{equation}
    E_\text{total}=\sqrt{1-\frac{2M}{r}}E^{\text{cm}}_0+E_0^\text{In}e^{-2n\pi/s_0}
\end{equation}

\begin{figure}[htbp]
    \centering
    \includegraphics[width=1\linewidth]{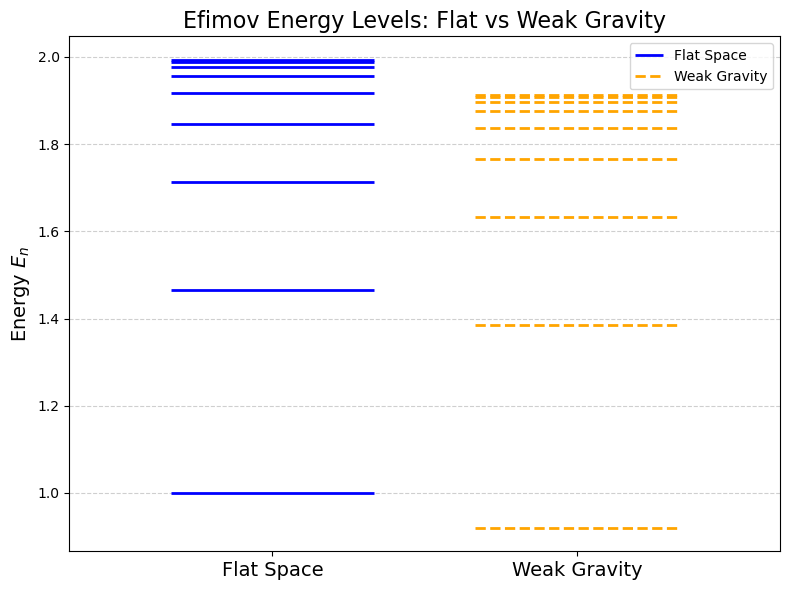}
    \caption{Total energy of the system is red-shifted $(2M/r=0.0392)$.}
    \label{f2}
\end{figure}

The figure \ref{f2} shows the Efimov energy level change under weak gravity influence.

\subsection{Efimov State in Strong Field}\label{strong field part}

The strong field is defined as: \textbf{the gravitational field is strong enough to alter the internal structure of the Efimov state.} This means that the Jacob coordinates can not be used since a vector can not be defined in a curved space, but the Jacob coordinates are the key to finding the Efimov state. This implies that a theoretical equilibrium must be established to include the gravitational effect in the internal structure, and the Jacob coordinate is able to be used.

By considering the Hamiltonian under Riemann Normal Coordinates (RNC) \cite{RNC} \cite{Parker}, where the gravitational perturbation is considered in the internal system. Generally, we need to expand the Laplacian operator in the curved space to include the curvature term. Firstly, applying the linear perturbation:

\begin{equation}
    g_{\mu\nu}=\delta_{\mu\nu}+h_{\mu\nu}+\mathcal{O}(h^2)
\end{equation}

Consider the Taylor expansion for the metric tensor:

\begin{eqnarray}
    g_{\mu\nu}(x)&=&\nonumber g_{\mu\nu}(x_0)+(\nabla_\alpha g_{\mu\nu})|_{x=x_0}(x^\alpha-x^0)\\&+&\frac{1}{2}(\nabla_\alpha \nabla_\beta g_{\mu\nu})|_{x=x_0}(x^\alpha-x^0)(x^\beta-x^0)
\end{eqnarray}

The covariant derivative for the metric tensor is zero, so that the second term vanishes. If we set the $x_0$ at the origin. We have:

\begin{equation}
    g_{\mu\nu}(0)=\delta_{\mu\nu}
\end{equation}

In the Riemann normal coordinates, the (0,4) Riemann curvature tensor is \cite{GR}:

\begin{equation}
R_{\mu\alpha\nu\beta} = \frac{1}{2} \left(
\partial_\alpha \partial_\beta g_{\mu\nu}
+ \partial_\mu \partial_\nu g_{\alpha\beta}
- \partial_\alpha \partial_\nu g_{\mu\beta}
- \partial_\mu \partial_\beta g_{\alpha\nu}
\right)
\end{equation}

 Solve it out and apply symmetries:

\begin{equation}
\nabla_\alpha\nabla_\beta g_{\mu\nu}=\partial_\alpha \partial_\beta g_{\mu\nu}
= -\frac{1}{3} \left( R_{\mu\alpha\nu\beta} + R_{\nu\alpha\mu\beta} \right)=-\frac{2}{3} R_{\mu\alpha\nu\beta}
\end{equation}

Finally, we get the approximation for the metric tensor:

\begin{equation}
g_{\mu\nu}(x) \approx \delta_{\mu\nu} - \frac{1}{3} R_{\mu\alpha\nu\beta} x^\alpha x^\beta
\end{equation}

Take it back to the Laplacian operator under this assumption. The Laplacian operator is:

\begin{equation}
    g^{\mu\nu}\nabla_\mu\nabla_\nu=\frac{1}{\sqrt{|g|}} \partial_\mu \left( \sqrt{|g|} g^{\mu\nu} \partial_\nu\right)
\end{equation}

After a series of heavy calculations, as shown in Appendix \ref{A}, we get:

\begin{equation}
    g^{\mu\nu}\nabla_\mu\nabla_\nu=\nabla^2-\frac{1}{3} R_\alpha^{\nu}x^\alpha\partial_\nu
\end{equation}

Therefore, the structure of the three identical Bosons in the strong gravitational field has changed:

\begin{equation}
\hat{H}\Psi=\hat{H}_0\Psi+\Delta\hat{H}\Psi=\sum_{i=1}^3\left[-\frac{\hbar^2}{2m_i} \nabla_{\boldsymbol{i}}^2 
    \Psi+\frac{\hbar^2}{6m_i} \left(R_\alpha^{\nu}x^\alpha\partial_\nu\right)_i\Psi\right]
\end{equation}

Where index $i$ means the degree of freedom in the Jacob matrix.

The Ricci tensor is written as:

\begin{equation}
    R_{\mu}^{\nu}(r,\theta)=\begin{pmatrix}
\frac{2M}{r^2(2M-r)} & 0 & 0 \\
0 & \frac{M}{r} & 0\\
0 & 0 & \frac{M}{r}\sin^2(\theta)
\end{pmatrix}
\end{equation}

The final expressions for the corrected Hamiltonian for a single particle may be written as:

\begin{equation}
\Delta \hat{H} = \frac{\hbar^2}{6m} \left[ 
R^1_1r\frac{\partial}{\partial r}+R^2_2\theta\frac{\partial}{\partial\theta} + 
R^3_3\phi\frac{\partial}{\partial\phi} 
\right]
\end{equation}

Once we considered the higher-order correction, the wave equation would be a highly nonlinear system. However, we can mainly focus on the lowest energy Boson system to reduce the complexity.

For the $s$-wave solution, the wavefunction is angular independent:

\begin{equation}
\Psi(r,\theta,\phi)=\Psi(r) \quad\text{with}\quad\hat{H}=-\frac{\hbar^2}{2m}\nabla^2+\frac{\hbar^2}{6m}
R^1_1r\frac{\partial}{\partial r}
\end{equation}

Now, for the three-particle system in Jacobi coordinates, we had discussed the centre mass wave solution in section \ref{Weakfield}, for the internal system ($\rho$,$\lambda$), it is reasonable to assume that the curvatures at the three Bosons positions are the same $\left(R^\nu_\alpha(r_1)=R^\nu_\alpha(r_2)=R^\nu_\alpha(r_3)\right)$ because the scale of the Efimov system is much smaller than the scale of the system to the blackhole. Therefore, the system can be treated as a whole, and the Ricci tensors are:

\begin{equation}
    R^\nu_\alpha(r_1)=R^\nu_\alpha(r_2)=R^\nu_\alpha(r_3)=R^\nu_\alpha(r_0)
\end{equation}

Where $r_0$ is the distance from the centre of mass of the system to the blackhole. The Hamiltonian can be written as:

\begin{equation}
\hat{H}=\sum_{i=1}^3-\frac{\hbar^2}{2m_i}\nabla^2_i+\frac{\hbar^2}{6}R^1_1(r_0)\sum_{i=1}^3\frac{r_i}{m_i}\frac{\partial}{\partial r_i}\label{49}
\end{equation}

Given that $m_1=m_2=m_3=m_0$, the Hamiltonian operator in the Jacobi operator is:

\begin{eqnarray}
    \hat{H}&=&-\frac{\hbar^2}{2m_0}\left(2\nabla_X^2+\frac{3}{2}\nabla_Y^2+\frac{1}{3}\nabla^2_{\text{cm}}\right)\\&+&\frac{M\hbar^2}{3m_0r_0^2(2M-r_0)}\left(X\frac{\partial}{\partial X}+Y\frac{\partial}{\partial Y}+r_0\frac{\partial}{\partial r_0}\right)
\end{eqnarray}

Rescale the equation above by introducing $\vec\rho=\frac{1}{\sqrt{2}}\vec X$ and $\vec\rho=\sqrt\frac{2}{3}\vec Y$:

\begin{eqnarray}
    \hat{H}&=&-\frac{\hbar^2}{2m_0}\left(\nabla_\rho^2+\nabla_\lambda^2+\frac{1}{3}\nabla^2_{\text{cm}}\right)\\&+&\frac{M\hbar^2}{3m_0r_0^2(2M-r_0)}(\rho\partial_\rho+\lambda\partial_\lambda+r_0\partial_{r_0})
\end{eqnarray}

Express it in the hyperspherical coordinate:
\begin{equation}
    \hat{H}_{\text{In}}=-\frac{\hbar^2}{2m_0}\left(\partial_\zeta^2+\frac{1}{\zeta}\partial_\zeta+\frac{1}{\zeta^2}\partial_\alpha^2\right)+\frac{M\hbar^2}{3m_0r_0^2(2M-r_0)}\zeta\partial_\zeta\label{54}
\end{equation}

The centre of mass equation is ignored since it has been discussed in the weak field part. The equation above shows that the hyper-radial equation of three identical bosons has changed in the strong field.

\section{Efimov Energy Level in Curved Space (Schwarzschild spacetime)}

This section focuses on the solution of the hyper-radial equation.  According to Eq.\ref{54}, we know that the hyper-angular equation stays the same. Therefore, the eigenvalue for it is still $s_0\approx1.00624i$. The Hamiltonian equation for the internal energy is:

\begin{equation}
    \left(\partial_\zeta^2+\frac{1}{\zeta}\partial_\zeta+\frac{s_0^2}{\zeta^2}-\frac{2M}{3r_0^2(2M-r_0)}\zeta\partial_\zeta\right)\psi=-\frac{2m_0E}{\hbar^2}\psi\label{gravdfe}
\end{equation}

This equation is hard to solve analytically. However, we can analyse it numerically, introducing the variable $x=\ln(\zeta)$. The equation \ref{gravdfe} can be rewritten into:

\begin{equation}
    \frac{\partial^2}{\partial x^2}\psi(x)+s_0^2\psi(x)+\epsilon e^{2x}\frac{\partial}{\partial x}\psi(x)=-\frac{2m_0E}{\hbar^2}\psi(x)\label{xequation}
\end{equation}

Where $\epsilon=\frac{2M}{3r_0^2(2M-r_0)}$, assume that the quantum system is near the Schwarzschild radius ($r_0\sim R_s=2M$). We consider the case of a low-mass black hole, assuming its mass is 20 times the solar mass ($20M_\odot$) \cite{blackhole}. The corresponding parameter $\epsilon$ would then be approximately $\epsilon\sim5\times10^{-11}$. As we have observed, even for currently detectable low-mass black holes, their perturbation to Efimov states remains extremely weak. Hence, it is reasonable to treat gravitational correction as the first-order perturbation.

\begin{figure}[htbp]
    \centering
    \includegraphics[width=1\linewidth]{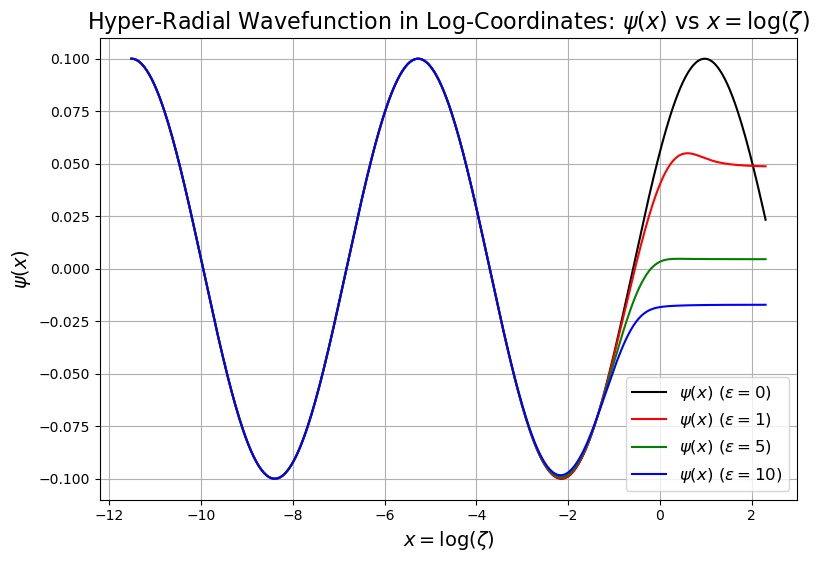}
    \caption{Comparison between wavefunction in flat space ($\epsilon=0$) and Schwarzschild space ($\epsilon=1,5,10$).}
    \label{3}
\end{figure}

Solving the equation numerically at the low energy limit $E\to0$, the wavefunction is shown in Figure \ref{3}.

It is clear that the gravitational effect is only effective at relatively long distances. From Eq. \ref{xequation}, since the exponential function can be approximated as zero at short distances, the solution reduces to a logarithmic oscillating form. This aligns with physical intuition: at sufficiently small scales, spacetime asymptotically approaches flatness ($g_{\mu\nu}\to\eta_{\mu\nu}$). This can also be seen as a verification of our calculation.

We can treat the fourth term as a perturbative term to the original system. Therefore, we can apply first-order perturbation theory to analyse the problem. The energy correction is:

\begin{equation}
    \Delta E^{(1)}_n=\frac{\bra{\Psi_n}\Delta\hat{H}\ket{\Psi_n}}{\braket{\Psi_n|\Psi_n}}=\frac{\iint\Psi_n^*\Delta\hat{H}\Psi_n\zeta d\zeta d\alpha}{\braket{\Psi_n|\Psi_n}}
\end{equation}

Where $\Psi_n$ is the original wavefunction under $\hat{H_0}$. We need to notice that the hyperspherical radial part solution is valid only in the low-energy limit. The solution is not well-defined in the whole Hilbert space, which means we cannot perform integration from $0$ to $\infty$.
Think carefully, this is not strange: although the Efimov effect predicts the existence of an infinite number of bound states in the zero-range and low-energy limit, it does not imply that the physical Efimov states extend over the entire real space without bound \cite{Braaten}. Instead, any realistic three-body system must involve an effective short-range cutoff $r_{\text{min}} \sim r_0$, set by the breakdown of the zero-range approximation, and a long-range cutoff $r_{\text{max}} \sim a$ \cite{1973}, determined by the scattering length or other external constraints.

Thus, the physical domain of the Efimov radial wave function is effectively limited to the interval:

$$
\zeta \in [\zeta_{\text{min}}, \zeta_{\text{max}}]
$$

which corresponds to a logarithmic window within which the discrete scale invariance is approximately valid. Beyond this window, either the zero-range assumption fails (at short distances) or the state merges into the continuum (at large distances). Therefore, when evaluating quantities such as energy corrections via perturbation theory, the integration must be restricted to this finite window.

In the discussion below, we chose the finite window for $\zeta$ from $\zeta_0$ to $\zeta_0e^ {q\pi/s_0}$, where $q\pi$ is the phase corresponding to the $\zeta_{\text{max}}$, to analyse energy correction at different energy levels $n$. Normalisation constant at one period is:

\begin{equation}
    \braket{F_n|F_n}=\frac{e^{2q\pi/s_0}(1+s_0^2-\cos[2q\pi]-s_0\sin[2q\pi])-s_0^2}{4(1+s_0^2)}\zeta_0^2
\end{equation}

Energy correction is given by:
\begin{equation*}
    \braket{F_n|\Delta\hat{H}|F_n}=-\frac{\hbar^2\epsilon}{2m_0}\int_{\zeta_0}^{\zeta_q}\int_0^{\frac{\pi}{2}}\Psi^*_n(\zeta,\alpha)\zeta^2\partial_\zeta\Psi_n(\zeta,\alpha)d\zeta d\alpha
\end{equation*}
\begin{equation}
    \braket{F_n|\Delta\hat{H}|F_n}=-\frac{\hbar^2\epsilon}{2m_0}\int_{\zeta_0}^{\zeta_q}F_n(\zeta)\zeta^2F'_n(\zeta)d\zeta\int_0^{\frac{\pi}{2}}|\Phi(\alpha)|^2d\alpha
\end{equation}

We have defined that $\zeta_q=\zeta_0e^{q\pi/s_0}$. The hyper-angular part is also normalised. Introducing a variable $x=s_0\ln(\zeta/\zeta_q)$ to rewrite the equation above:

\begin{eqnarray}
    \Delta E_n^{(1)}&=&\frac{\hbar^2\epsilon}{2m_0}\frac{\zeta_0^2}{\braket{F_n|F_n}}\int_0^{q\pi}e^{2x/s_0}\sin x\cos xdx\nonumber\\&=&A\frac{\hbar^2M}{3m_0r_0^2(2M-r_0)}\label{Firstorder}
\end{eqnarray}

Where:

\begin{equation}
    A=\frac{s_0^2+s_0e^{2q\pi/s_0}(\sin[2q\pi]-s_0\cos[2q\pi])}{s_0^2-e^{2q\pi/s_0}(1+s_0^2-\cos[2q\pi]-s_0\sin[2q\pi])}
\end{equation} 

The result is independent to the quantum number $n$ and the short-range cutoff $\zeta_0$, which indicates that the gravitational field will not change the internal energy structure to first-order accuracy. In SI units, the Efimov energy in the Schwarzschild spacetime is given by:

\begin{equation}
    E_n=\sqrt{1-\frac{2GM}{c^2r_0}}E^{\text{cm}}_0+E^{\text{In}}_0e^{-2n\pi/s_0}+A\frac{\hbar^2GM}{3m_0r_0^2(2GM-c^2r_0)}\label{62}
\end{equation}

\begin{figure}[htbp]
    \centering
    \includegraphics[width=1\linewidth]{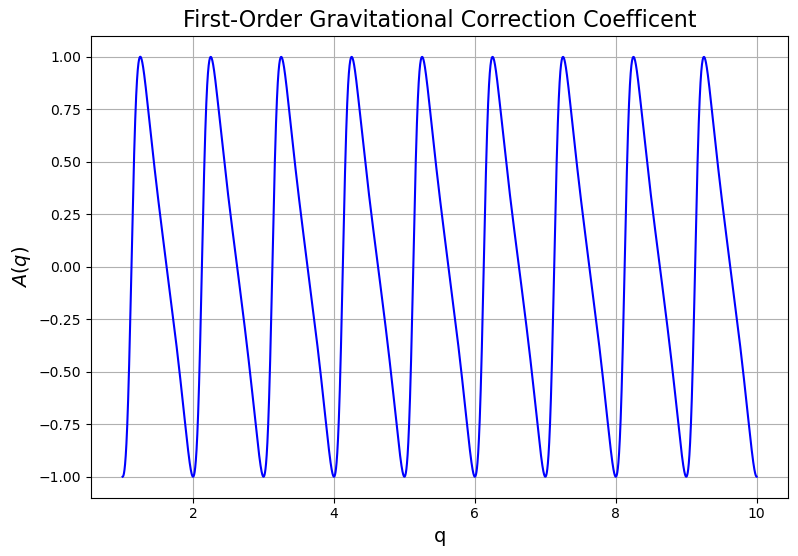}
    \caption{The gravitational energy correction is periodic under different length scales.}
    \label{4}
\end{figure}

Where $E_0^{\text{In}}=-\hbar^2/2m_0\zeta_0^2$. $A$ can be treated as a function of scattering length with parameter $q$, as \ref{4} shows.

The gravitational correction to the Efimov system is too weak, with an energy shift of $\Delta E^{(1)} \sim 10^{-31}\text{eV}$ for a source mass of $M=20M_\odot$.

\section{Experimental Possibilities and Future Directions}

The experimental validation of this theory faces two primary challenges: firstly, the absence of strong gravitational sources analogous to black holes in laboratory settings, and secondly, the extremely weak perturbations induced by gravitational effects.

Although it is impossible to directly create curved spacetime as described by general relativity in the lab, its physical effects can be simulated using analogue models. In 1981, Unruh first proposed that transonic flow could be used to simulate black hole horizons and their quantum effects, establishing the first analogue model of black holes \cite{Unruh}. Later in 2016, Jeff Steinhauer successfully observed quantum entanglement in acoustic Hawking radiation using a Bose-Einstein condensate (BEC) system, experimentally validating the feasibility of this approach. \cite{Hawking}

The validation of our theory could likewise adopt the aforementioned approach by simulating curved spacetime in laboratory settings. For instance, by leveraging the quantum behaviour of ultracold atoms in periodic optical lattices and manipulating laser parameters (e.g., intensity, frequency, polarisation, hopping parameters), an equivalent "curved spacetime metric" can be engineered. This allows the simulation of quantum field dynamics in a curved background. \cite{lab}

Building on established analogue gravity platforms, our future work proposes to experimentally realise a three-dimensional Schwarzschild metric using ultracold atoms in optical lattices while simultaneously implementing an Efimov trimer system \cite{lab}. For weak-field validation, we will configure the metric perturbation to small values ($g_{mn} \approx \eta_{mn} + \epsilon h_{mn} \quad\text{with} \quad|\epsilon|\ll1$) and measure the characteristic decrease in atomic number density to reconstruct Efimov energy levels \cite{Braaten}, thereby testing whether the observed level redshifts conform to weak-field predictions. In the strong-field regime, we will exploit the quantitative relationship between scattering parameter $q$ and coefficient $A$ (as shown in Fig. \ref{4}) to systematically vary the scattering length while monitoring the Efimov spectrum for periodic oscillations - a signature effect predicted by our theoretical framework under strong field perturbations. This approach will provide comprehensive experimental verification of how curved spacetime geometry modulates quantum few-body states across different gravitational field strengths.

\section{Conclusion}

In the Jacobi coordinate system, the derivation of Efimov states in flat spacetime is reproduced. For the Schwarzschild background, the Efimov system is decoupled into centre-of-mass and internal structures, obtaining the centre-of-mass energy expression in weak gravitational fields by combining the Schrödinger equation under general relativity.

Furthermore, through a Riemann normal coordinate expansion of the Hamiltonian operator—while preserving coordinate flatness—the Hamiltonian correction for strong gravitational fields is derived. The analysis reveals that for three bosons of equal mass, gravitational perturbations only manifest in the hyperradial equation, implying no change in the relative positions of Efimov states. However, if the boson masses differ, the perturbation contributes to the hyperangular operator, ultimately modifying the hyperangular equation. This could alter the distribution of Efimov energy levels. Using first-order perturbation theory, we quantified the gravitational corrections to these levels (see Eq. \ref{Firstorder} for details).

Although this work theoretically extends the discussion of Efimov states from flat spacetime to gravitational backgrounds, cases with unequal boson masses were not discussed. Additionally, the experimental verification of these theoretical predictions could be pursued through analogue gravity systems, where ultra-cold atoms in precisely controlled optical potentials may simulate the equivalent spacetime curvature effects.

\begin{acknowledgments}
I am deeply grateful to my mother, Shuping Yang, for her constant emotional support and patient understanding during the challenging phases of this research. Her unwavering encouragement proved invaluable to the completion of this work.
\end{acknowledgments}

\appendix

\section{Derivation of the Hamiltonian Operator in Strong Fields}\label{A}

In this part, we will provide a detailed derivation of the corrected Hamiltonian Operator in strong gravitational fields.

What we are going to do is consider the gravitational perturbation in the internal system. Generally, we need to expand the Laplacian operator in the curved space to include the curvature term. Firstly, applying the linear perturbation:

\begin{equation}
    g_{\mu\nu}=\delta_{\mu\nu}+h_{\mu\nu}+\mathcal{O}(h^2)
\end{equation}

Consider the Taylor expansion for the metric tensor:

\begin{eqnarray}
    g_{\mu\nu}(x)&=&g_{\mu\nu}(x_0)+(\nabla_\alpha g_{\mu\nu})|_{x=x_0}(x^\alpha-x^0)\nonumber\\&+&\frac{1}{2}(\nabla_\alpha \nabla_\beta g_{\mu\nu})|_{x=x_0}(x^\alpha-x^0)(x^\beta-x^0)+...
\end{eqnarray}

The covariant derivative for the metric tensor is zero, so that the second term vanishes. If we set the $x_0$ at the origin. We have:

\begin{equation}
    g_{\mu\nu}(0)=\delta_{\mu\nu}
\end{equation}

In the Riemann normal coordinates, the (0,4) Riemann curvature tensor is:

\begin{equation}
R_{\mu\alpha\nu\beta} = \frac{1}{2} \left(
\partial_\alpha \partial_\beta g_{\mu\nu}
+ \partial_\mu \partial_\nu g_{\alpha\beta}
- \partial_\alpha \partial_\nu g_{\mu\beta}
- \partial_\mu \partial_\beta g_{\alpha\nu}
\right)
\end{equation}

 Solve it out and apply Symmetries:

\begin{equation}
\nabla_\alpha\nabla_\beta g_{\mu\nu}=\partial_\alpha \partial_\beta g_{\mu\nu}
= -\frac{1}{3} \left( R_{\mu\alpha\nu\beta} + R_{\nu\alpha\mu\beta} \right)=-\frac{2}{3} R_{\mu\alpha\nu\beta}
\end{equation}

Finally, we get the approximation for the metric tensor:

\begin{equation}
g_{\mu\nu}(x) \approx \delta_{\mu\nu} - \frac{1}{3} R_{\mu\alpha\nu\beta} x^\alpha x^\beta
\end{equation}

The good thing about the approximation is: we can applying it to the kinetic energy in Hyperspherical Coordinates as shown in equation (7).\\

Next, we need to find an expression for the Laplacian operator under this assumption. The Laplacian operator is:

\begin{equation}
    g^{\mu\nu}\nabla_\mu\nabla_\nu=\frac{1}{\sqrt{|g|}} \partial_\mu \left( \sqrt{|g|} g^{\mu\nu} \partial_\nu\right)
\end{equation}

The determinant for the metric can be approximated by linear perturbation theory:

\begin{equation}
    \det(g_{\mu\nu})=\det(\delta_{\mu\nu}+h_{\mu\nu})=1+\text{Tr}(h)+\mathcal{O}(h^2)
\end{equation}

Which means:

\begin{equation}
    \sqrt{|g|}=1+\frac{1}{2}\text{Tr}(h)=1-\frac{1}{6}\delta^{\mu\nu}R_{\mu\alpha\nu\beta}x^\alpha x^\beta=1-\frac{1}{6}R_{\alpha\beta}x^\alpha x^\beta
\end{equation}

and:

\begin{equation}
    \frac{1}{\sqrt{|g|}}=1-\frac{1}{2}\text{Tr}(h)=1+\frac{1}{6}\delta^{\mu\nu}R_{\mu\alpha\nu\beta}x^\alpha x^\beta=1+\frac{1}{6}R_{\alpha\beta}x^\alpha x^\beta
\end{equation}

For the inverse matric tensor, it is given by:

\begin{equation}
    g^{\mu\nu}=\delta^{\mu\nu} + \frac{1}{3} R_{\alpha\beta}^{\mu\nu} x^\alpha x^\beta
\end{equation}

Where:

\begin{equation}
    R_{\alpha\beta}^{\mu\nu}=\delta^{\mu\rho}\delta^{\nu\sigma}R_{\rho\alpha\sigma\beta}=R_{\mu\alpha\nu\beta}
\end{equation}

Therefore:

\begin{equation}
    A^{\mu\nu}=\left(1-\frac{1}{6}R_{\alpha\beta}x^\alpha x^\beta\right)\left(\delta^{\mu\nu} + \frac{1}{3} R_{\alpha\beta}^{\mu\nu} x^\alpha x^\beta\right)
\end{equation}

Ignore the higher-order term:

\begin{equation}
    A^{\mu\nu}=\delta^{\mu\nu}+\left(\frac{1}{3} R_{\alpha\beta}^{\mu\nu}-\frac{1}{6}\delta^{\mu\nu}R_{\alpha\beta}\right)x^\alpha x^\beta+\mathcal{O}(x^3)
\end{equation}

Next, we shall calculate the expression for the Laplacian operator under linear perturbation:

\begin{eqnarray}
    \partial_\mu\left(A^{\mu\nu}\partial_\nu\right)&=&\partial_\mu A^{\mu\nu}\partial_\nu+A^{\mu\nu}\partial_\mu\partial_\nu\\&=&\left(\frac{1}{3} R_{\alpha\beta}^{\mu\nu}-\frac{1}{6}\delta^{\mu\nu}R_{\alpha\beta}\right)\left(\delta^\alpha_\mu x^\beta+x^\alpha\delta^\beta_\mu\right)\partial_\nu\nonumber\\&&+A^{\mu\nu}\partial_\mu\partial_\nu
\end{eqnarray}

Simplify the first term:

\begin{eqnarray}
    \partial_\mu\left(A^{\mu\nu}\partial_\nu\right)\nonumber&=&\left(\frac{1}{3} R_{\mu\beta}^{\mu\nu}-\frac{1}{6}\delta^{\mu\nu}R_{\mu\beta}\right) x^\beta\partial_\nu\nonumber\\&+&\left(\frac{1}{3} R_{\alpha\mu}^{\mu\nu}-\frac{1}{6}\delta^{\mu\nu}R_{\alpha\mu}\right) x^\alpha\partial_\nu\nonumber\\&+&A^{\mu\nu}\partial_\mu\partial_\nu
\end{eqnarray}

Noticed that:

\begin{equation*}
    R_{\alpha\mu}^{\mu\nu}=-R_{\mu\alpha}^{\mu\nu}\quad \text{and}\quad R_{\alpha\mu}=R_{\mu\alpha}
\end{equation*}

Therefore:

\begin{eqnarray}
    \partial_\mu\left(A^{\mu\nu}\partial_\nu\right)&=&\left[\delta^{\mu\nu}+\left(\frac{1}{3} R_{\alpha\beta}^{\mu\nu}-\frac{1}{6}\delta^{\mu\nu}R_{\alpha\beta}\right)x^\alpha x^\beta\right]\partial_\mu\partial_\nu\nonumber\\&-&\frac{1}{3} \delta^{\mu\nu}R_{\mu\alpha}x^\alpha\partial_\nu
\end{eqnarray}

\begin{eqnarray}
    \partial_\mu\left(A^{\mu\nu}\partial_\nu\right)&=&\delta^{\mu\nu}\partial_\mu\partial_\nu-\frac{1}{3} \delta^{\mu\nu}R_{\mu\alpha}x^\alpha\partial_\nu\\&+&\left(\frac{1}{3} R_{\alpha\beta}^{\mu\nu}-\frac{1}{6}\delta^{\mu\nu}R_{\alpha\beta}\right)x^\alpha x^\beta\partial_\mu\partial_\nu
\end{eqnarray}

Finally, we can get:

\begin{eqnarray}
    g^{\mu\nu}\nabla_\mu\nabla_\nu&=&\left[1+\frac{1}{6}R_{\alpha\beta}x^\alpha x^\beta\right]\Bigg[\delta^{\mu\nu}\partial_\mu\partial_\nu-\frac{1}{3} R_\alpha^{\nu}x^\alpha\partial_\nu\nonumber\\&+&\left(\frac{1}{3} R_{\alpha\beta}^{\mu\nu}-\frac{1}{6}\delta^{\mu\nu}R_{\alpha\beta}\right)x^\alpha x^\beta\partial_\mu\partial_\nu\Bigg]
\end{eqnarray}

Where $R^\nu_\alpha=\delta^{\mu\nu}R_{\mu\alpha}$, we will mainly focus on the contribution from the second term (e.g. the curvature correction), which means we can ignore higher order terms ($\mathcal{O}(x^3)$):

\begin{eqnarray}
    g^{\mu\nu}\nabla_\mu\nabla_\nu&=&\delta^{\mu\nu}\partial_\mu\partial_\nu-\frac{1}{3} R_\alpha^{\nu}x^\alpha\partial_\nu\nonumber\\&+&\left(\frac{1}{3} R_{\alpha\beta}^{\mu\nu}-\frac{1}{6}\delta^{\mu\nu}R_{\alpha\beta}\right)x^\alpha x^\beta\partial_\mu\partial_\nu\nonumber\\&+&\frac{1}{6}R_{\alpha\beta}x^\alpha x^\beta\delta^{\mu\nu}\partial_\mu\partial_\nu
\end{eqnarray}

Simplify it to get:

\begin{equation}
    g^{\mu\nu}\nabla_\mu\nabla_\nu=\delta^{\mu\nu}\partial_\mu\partial_\nu-\frac{1}{3} R_\alpha^{\nu}x^\alpha\partial_\nu+\frac{1}{3} R_{\alpha\beta}^{\mu\nu}x^\alpha x^\beta\partial_\mu\partial_\nu
\end{equation}

Because of the antisymmetry of the Riemann tensor, the third term is identically zero:

\begin{equation}
    g^{\mu\nu}\nabla_\mu\nabla_\nu=\nabla^2-\frac{1}{3} R_\alpha^{\nu}x^\alpha\partial_\nu
\end{equation}

Therefore, the structure of the three identical Bosons in the strong gravitational field has changed:

\begin{eqnarray}
    \hat{H}\Psi&=&\hat{H}_0\Psi+\Delta\hat{H}\Psi\\&=&\sum_{i=1}^3\left[-\frac{\hbar^2}{2m_i} \nabla_{\boldsymbol{i}}^2 
    \Psi+\frac{\hbar^2}{6m_i} \left(R_\alpha^{\nu}x^\alpha\partial_\nu\right)_i\Psi\right]
\end{eqnarray}

Where index $i$ means the degree of freedom in the Jacob matrix.

The Ricci tensor in Schwarzschild space is written as:

\begin{equation}
    R_\mu^\nu=g^{\nu\beta}R_{\beta\mu}=g^{\nu\beta}R^\rho_{\beta\rho\mu}
\end{equation}

\begin{equation}
    R_{\beta\mu}=R^\rho_{\beta\rho\mu}=\partial_\rho\Gamma^\rho_{\mu\beta}-\partial_\mu\Gamma^\rho_{\rho\beta}+\Gamma^\rho_{\rho\lambda}\Gamma^\lambda_{\mu\beta}-\Gamma_{\mu\lambda}^\rho\Gamma^\lambda_{\rho\beta}
\end{equation}

The Christoffel connection $\Gamma_{\mu\nu}^\sigma$ is \cite{GR}:

\begin{equation}
    \Gamma^{\sigma}_{\mu\nu} = \frac{1}{2} g^{\sigma\rho} \left( \partial_{\mu} g_{\nu\rho} + \partial_{\nu} g_{\rho\mu} - \partial_{\rho} g_{\mu\nu} \right)
\end{equation}

Where $g^{\sigma\rho}$ is the inverse matrix of $g_{\sigma\rho}$, the all $g_{\sigma\rho}$ and $g^{\sigma\rho}$ components are given by:

\begin{eqnarray}
    g_{rr}=\frac{1}{1-\frac{2M}{r}}, \quad g_{\theta\theta}=r^2, \quad g_{\phi\phi}=r^2\sin^2\theta
\end{eqnarray}

\begin{eqnarray}
    g^{rr}=1-\frac{2M}{r}, \quad g^{\theta\theta}=1/r^2, \quad g^{\phi\phi}=\frac{1}{r^2\sin^2\theta}
\end{eqnarray}

Compute all non-vanishing Christoffel connections:

\begin{equation}
\begin{cases}
\Gamma^r_{rr}=\frac{M}{2Mr-r^2}\\
\Gamma^r_{\theta\theta}=2M-r\\
\Gamma^r_{\phi\phi}=(2M-r)\sin^2\theta
\end{cases}
\end{equation}

\begin{equation}
\begin{cases}
\Gamma^\theta_{\theta r}=\Gamma^\theta_{r\theta}=\frac{1}{r}\\
\Gamma^r_{\theta\theta}=2M-r\\
\Gamma^\theta_{\phi\phi}=-\frac{1}{2}\sin(2\theta)
\end{cases}
\end{equation}

\begin{equation}
\begin{cases}
\Gamma^\phi_{r\phi}=\Gamma^\phi_{\phi r}=\frac{1}{r}\\
\Gamma^\phi_{\theta\phi}=\Gamma^\phi_{\phi\theta}=\cot(\theta)
\end{cases}
\end{equation}

The non-vanishing Ricci tensor components are:

\begin{eqnarray}
    R_{rr}=\frac{2M}{r^2(2M-r)},R_{\theta\theta}=\frac{M}{r},R_{\phi\phi}=\frac{M\sin^2\theta}{r}
\end{eqnarray}

Therefore:

\begin{equation}
\begin{cases}
R^r_r=R_{rr}=\frac{2M}{r^2(2M-r)}\\
R^\theta_\theta=R_{\theta\theta}=\frac{M}{r}\\
R^\phi_\phi=R_{\phi\phi}=\frac{M}{r}\sin^2(\theta)
\end{cases}
\end{equation}

\begin{equation}
    R_{\mu}^{\nu}=\begin{pmatrix}
\frac{2M}{r^2(2M-r)} & 0 & 0 \\
0 & \frac{M}{r} & 0\\
0 & 0 & \frac{M}{r}\sin^2(\theta)
\end{pmatrix}
\end{equation}

\section{Hamiltonian Operator under Jacobi Coordinate}\label{B}

Firstly, the coordinate transformation between two coordinate systems is given by:

\begin{equation}
    \begin{pmatrix}
        X \\ Y \\ r_0
    \end{pmatrix}
    =\begin{pmatrix}
-1 & 1 & 0 \\
-\frac{m_1}{m_1+m_2} & -\frac{m_2}{m_1+m_2} & 1\\
\frac{m_1}{M} & \frac{m_2}{M} & \frac{m_3}{M}
\end{pmatrix}\begin{pmatrix}
    {x}_1 \\ {x}_2 \\
    {x}_3
\end{pmatrix}
\end{equation}

Where $X^\alpha=\{X,Y,r_0\}$ is Jacobi coordinate components and $x^i=\{x_1,x_2,x_3\}$ is normal Cartesian coordinate components.

The expression above can  be written into tensor form:

\begin{eqnarray}
    dX^\alpha=T^\alpha_id{x}^i\quad \text{and} \quad dx^i=T_\alpha^idX^\alpha
\end{eqnarray}

$T^\alpha_i$ and $T^i_\alpha$ are inverse matrices of each other. Consider the eq \ref{49}, the total Hamiltonian operator can be written in tensor form:

\begin{eqnarray}
    \hat H=-\frac{\hbar^2}{2m_i}\left(\frac{\partial}{\partial x^i}\right)^2+\frac{M\hbar^2}{3r_0^2(2M-r_0)}\frac{x^i}{m_i}\frac{\partial}{\partial{x^i}}
\end{eqnarray}

Applying the chain rule, we get the expression under the Jacobi coordinates:

\begin{eqnarray}
    \hat H&=&-\frac{\hbar^2}{2m_i}T_i^\alpha T_i^\beta\partial_\alpha\partial_\beta\nonumber\\&&+\frac{M\hbar^2}{3r_0^2(2M-r_0)}T^i_\beta T^\alpha_i\frac{X^\beta}{m_i}\partial_\alpha
\end{eqnarray}

We can write it in matrix form:

\begin{eqnarray}
    \hat H=-\frac{\hbar^2}{2}A^{\alpha\beta}\partial_\alpha\partial_\beta+\frac{M\hbar^2}{3r_0^2(2M-r_0)}B^\alpha_\beta X^\beta\partial_\alpha
\end{eqnarray}

where $\alpha$ denotes the row index and $\beta$ denotes the column index. Two matrices are defined as:

\begin{eqnarray}
    A^{\alpha\beta}=\frac{T_i^\alpha T_i^\beta}{m_i}\quad \text{and}\quad B^\alpha_\beta=\frac{T^i_\beta T^\alpha_i}{m_i}
\end{eqnarray}

\begin{eqnarray}
    A^{\alpha\beta}=\begin{pmatrix}
\frac{1}{m_1}+\frac{1}{m_2} & 0 & 0 \\
0 & \frac{1}{m_{12}}+\frac{1}{m_3} & 0\\
0 & 0 & \frac{1}{M}
\end{pmatrix}
\end{eqnarray}

\begin{eqnarray}
    B^\alpha_\beta=\begin{pmatrix}
\frac{m_1^2+m_2^2}{m_1m_2m_{12}} & \frac{(m_2-m_1)m_3}{m_1m_2M} & \frac{m_1-m_2}{m_1m_2} \\
\frac{m_2-m_1}{m_{12}^2} & \frac{m_{12}^2+2m_3^2}{m_{12}m_3M} & \frac{m_{12}-2m_3}{m_{12}m_3}\\
\frac{m_1-m_2}{m_{12}M} & \frac{m_{12}-2m_3}{M^2} & \frac{3}{M}
\end{pmatrix}
\end{eqnarray}
Where $m_{12}=m_1+m_2$ and $M=m_1+m_2+m_3$. The non-perturbative Hamiltonian operator is:

\begin{eqnarray}
    \hat{H}_0&=&-\frac{\hbar^2}{2}A^{\alpha\beta}\partial_\alpha\partial_\beta\\&=&-\frac{\hbar^2}{2\mu_X}\nabla_X^2-\frac{\hbar^2}{2\mu_Y}\nabla_Y^2-\frac{\hbar^2}{2M}\nabla_{\text{cm}}^2
\end{eqnarray}

where:

\begin{itemize}
  \item $\mu_X = \dfrac{m_1 m_2}{m_1 + m_2}$: reduced mass of particles 1 and 2
  \item $\mu_Y =\dfrac{m_3 (m_1 + m_2)}{m_1 + m_2 + m_3}$: reduced mass of particle 3 relative to the centre of mass of particles 1 and 2.
\end{itemize}

As for the matrix $B^\alpha_\beta$, the matrix will only be orthogonal when $m_1=m_2=m_3=m_0$, as previously discussed in part \ref{strong field part}. If the three masses are unequal, the off-diagonal elements will also contribute to the perturbative energy, making the calculations more complicated. For this reason, we restrict our discussion to systems of equal-mass bosons.

\end{document}